\documentclass[fdp,fleqn]{w-art}
\usepackage{times,cite}
\usepackage{amssymb,amsmath}

\DeclareMathOperator{\tr}{Tr}
\newcommand{\Hodge}{\mathop{\star}}

\newcommand{\e}{\textrm{e}}
\newcommand{\de}{\textrm{d}}

\newcommand{\cN}{\mathcal{N}}

\newcommand{\abs}[1]{\lvert #1\rvert}
\journalname{}{}{Typeset in Fortschr.~Phys.~style}

\begin{document}
\DOIsuffix{theDOIsuffix}
\pagespan{1}{}

\title[Black holes, first-order flow equations and geodesics]%
{Black holes, first-order flow equations \\and geodesics on symmetric spaces}

\author[J.~Perz]{Jan Perz\inst{1,}%
  \footnote{Corresponding author\quad E-mail:~\textsf{jan.perz@fys.kuleuven.be},
            Phone: +32\,16\,327248,
            Fax: +32\,16\,327986}}
\address[\inst{1}]{Afdeling Theoretische Fysica,
Katholieke Universiteit Leuven\\Celestijnenlaan 200D bus 2415, 3001 Heverlee, Belgium}
\author[P.~Smyth]{Paul Smyth\inst{2,}\footnote{E-mail:~\textsf{paul.smyth@desy.de}}}
\address[\inst{2}]{II. Institut f\"ur Theoretische Physik der
Universit\"at Hamburg\\Luruper Chaussee 149, 22761 Hamburg, Germany}
\author[T.~Van Riet]{Thomas Van Riet\inst{3,}\footnote{E-mail:~\textsf{thomas.vanriet@physics.uu.se}}}
\address[\inst{3}]{Institutionen f\"or fysik och astronomi,
Uppsala Universitet\\ Box 803, 751 08 Uppsala, Sweden}
\author[B.~Vercnocke]{Bert Vercnocke\inst{1,}\footnote{E-mail:~\textsf{bert.vercnocke@fys.kuleuven.be}}}

\begin{abstract}
For both extremal and non-extremal spherically symmetric black holes
in theories with massless scalars and vectors coupled to gravity, we
derive a general form of first-order gradient flow equations, equivalent to the equations of motion. For theories that have a
symmetric moduli space after a dimensional reduction over the timelike
direction, we discuss the condition for such a gradient flow to exist.

\noindent
This note reviews the results of \cite{Perz:2008kh}.
\end{abstract}

\begin{flushright}
KUL-TF-09/01, UUITP-03/09
\end{flushright}

\maketitle

\section{Introduction}

Black hole solutions to extensions of general relativity,
such as the various kinds of supergravity naturally occurring in the
low-energy effective description of superstrings, often exhibit
features unknown from pure Einstein's theory. One such feature,
distinctive for extremal black holes in gravity coupled to scalar and
vector fields, is the attractor phenomenon
\cite{Ferrara:1995ih,Ferrara:1996dd,Ferrara:1996um,Ferrara:1997tw}.
It causes the end-points of the radial evolution of the scalars -- their
values on the event horizon -- to be determined by the charges
associated with the vectors and to be insensitive to the values of the scalars
at spatial infinity. In particular, for extremal solutions that are
supersymmetric, the evolution is governed by first-order (BPS)
equations: the scalar fields follow a gradient flow in target space.
Recently it was noticed, however,
that non-supersymmetric, extremal black holes may also obey
first-order gradient flows \cite{Ceresole:2007wx}. Moreover, examples of
first-order equations have been found for some non-extremal
(and hence neither supersymmetric nor attractive) black holes
\cite{Lu:2003iv,Miller:2006ay,Janssen:2007rc,Cardoso:2008gm}.

Two main questions arise: What is the general form of first-order flow
equations for black holes? When does a gradient flow exist?
For extremal black holes, these questions were first addressed in
\cite{Ceresole:2007wx,Cardoso:2007ky,Andrianopoli:2007gt,
Ferrara:2008ap}, while \cite{Andrianopoli:2007gt} suggested in
addition a possible extension to non-extremal black holes. Our work
\cite{Perz:2008kh}, which we shall briefly review here,
offers a general answer, valid for static and spherically symmetric
solutions -- extremal and non-extremal alike.

In section \ref{s:FlowFormalism} we derive the generalised form of
first-order flow equations by demanding that the action be written as
a sum of perfect squares. The conditions for the scalar fields to obey
a first-order gradient flow are then found in section \ref{s:Existence}.
In section \ref{s:FreeParticle} we analyse the case when the theory
after dimensional reduction over time describes a non-linear sigma model
on a symmetric space. We end with a discussion of some explicit examples
in four and five dimensions in section \ref{s:Discussion}.

\section{Effective action and flow equations}\label{s:FlowFormalism}

Consider gravity coupled to a number of neutral scalars $\phi^a$ and
vector fields $A^I$ in $D+1$ dimensions,
\begin{equation}\label{E-M-D-action}
S = \int \de^{D+1}x \sqrt{|g|}\Bigl(\mathcal{R}_{D+1} -
\tfrac{1}{2}G_{ab}\partial_\mu\phi^a\partial^\mu\phi^b -
\tfrac{1}{2}\mu_{IJ} F^I_{\mu\nu} F^{J\,\mu\nu}\Bigr),
\end{equation}
where $G_{ab}$ and $\mu_{IJ}$ are functions that depend on the
scalars $\phi ^ a$, and $F^I_{\mu\nu}$ are Abelian field
strengths.\footnote{When $D+1=4$, there can be another term  of the
form $-\tfrac12\nu_{IJ}(\phi)F^I_{\mu\nu}(\Hodge F^J)^{\mu\nu}$ in the
action.} Greek indices are raised and lowered with the spacetime
metric $g_{\mu\nu}$ and $g=\det g_{\mu\nu}$. The most general metric
describing static, spherically symmetric black hole solutions of the
theory described by the action (\ref{E-M-D-action}) is
\begin{equation}
\de s^2_{D+1} =  - \e^{2\beta\varphi}\de t^2
+ \e^{2\alpha\varphi} \Bigl(\e^{2(D-1)A}\de\tau^2
+ \e^{2A}\de\Omega_{D-1}^2\Bigr),\quad
\e^{-(D-2)A}=\gamma^{-1}\sinh[(D-2)\gamma\tau]\,,
\label{metric}
\end{equation}
where $\alpha=-1/\sqrt{2(D-1)(D-2)}$, $\beta=-(D-2)\alpha$, $\gamma$
is a constant, and the scalars depend solely on the radial coordinate:
$\varphi =
\varphi(\tau)$, $\phi^a = \phi^a(\tau)$.
The equations of motion for scalar fields of this system can be
derived from a one-dimensional effective action
\begin{equation}\label{a1}
S=\int\de\tau\,\,\Bigl(
-\tfrac{1}{2}\dot{\varphi}^2 -
\tfrac{1}{2}G_{ab}\dot{\phi}^a\dot{\phi}^b -
\e^{2\beta\varphi}V(\phi^a)\Bigr),
\end{equation}
where a dot denotes a derivative with respect to $\tau$ and the
effective potential $V(\phi^a)$
results from solving for the vector fields in terms of the charges.
This action is supplemented with a Hamiltonian
constraint, which states that the radial evolution of
the fields happens on a slice of constant total energy
\begin{equation}
(D-1)(D-2)\gamma^2=\tfrac12\dot{\varphi}^2+
\tfrac12 G_{ab}\dot{\phi}^a \dot{\phi}^b -
\e^{2\beta\varphi}V(\phi)\equiv
E\,.\label{HamiltonianI}
\end{equation}
The constraint is the remnant of the original $D+1$-dimensional
Einstein equations that is not reproduced by the effective action
(\ref{a1}).

This type of effective action was first introduced in the context of
supersymmetric black holes in $\cN=2$ supergravity in four
dimensions \cite{Gibbons:1996af,Ferrara:1997tw},
where it was also observed that the black hole potential
can be derived from a superpotential, proportional to the modulus of
the central charge $\abs{Z}$
\begin{equation}
V = \tfrac{1}{2}\beta^2 W^2 + \tfrac{1}{2}G^{ab}\partial_a W
\partial_b W\,.
\end{equation}
The terms in the effective action can then be rearranged as a sum of
squares of the BPS equations, in this setting known as the attractor
flow equations:
\begin{align}
\dot{\varphi}&=\pm\beta\e^{\beta\varphi}W\label{FLOW'II}\,,\\
\dot{\phi}^a&=\pm\e^{\beta\varphi}\partial^{a}W\label{FLOW'III}\,.
\end{align}

Only relatively recently has it been noticed \cite{Ceresole:2007wx}
that the above rewriting is not unique, and when $W\not\propto\abs{Z}$ can
describe extremal black holes that are not supersymmetric. It has
also been demonstrated by \cite{Cardoso:2007ky}
(and corroborated by \cite{Ferrara:2008ap}) that when these
four-dimensional equations are viewed as dimensionally reduced
five-dimensional flows, the ambiguity in defining $W$ is not
merely a residue of supersymmetry in one dimension higher.
An ansatz for $W$ reproducing all the known black hole attractors in
$\cN > 2$ supergravities in $D + 1 = 4$ has been constructed in
\cite{Andrianopoli:2007gt}.

Andrianopoli et al.\ \cite{Andrianopoli:2007gt} explored also the possibility of
formulating a superpotential capable of describing both extremal
(supersymmetric and non-supersymmetric) as well as non-extremal
black holes. The required generalization of
the superpotential $W$ would consist in adding an explicit dependence
on the radial parameter $\tau$. Here we outline a different
approach, presented and exemplified in \cite{Perz:2008kh}.

Assuming that there exists a `generalised superpotential'
$Y(\varphi,\phi^a)$, such that
\begin{equation}
e^{2\beta\varphi}V(\phi^a)=\tfrac{1}{2}\partial_{\varphi}
Y\partial_{\varphi} Y
+\tfrac{1}{2} \partial_a Y\partial^a Y + \Delta\,,\label{Y-equation}
\end{equation}
where $\Delta$ is a constant, the effective action (\ref{a1}) can
be written in the form\footnote{Without loss of generality
(redefinition of variables) we choose the plus sign within the
squares.}
\begin{align}
S=&-\frac{1}{2}
\int\de\tau\,\Bigl[(\dot{\varphi}+\partial_\varphi Y)^2 +
(\dot{\phi}^a +\partial^a Y)^2\Bigr],\label{eq:SumSquares1}
\end{align}
plus a total derivative. We will address the question of
existence in the next section.

Demanding a stationary point of the action produces \emph{generalised
flow equations}
\begin{align}
\dot{\varphi}+\partial_{\varphi}Y&=0\label{FLOWII}\,,\\
\dot{\phi}^a + G^{ab}\partial_{b}Y&=0\label{FLOWIII}\,.
\end{align}
The Hamiltonian constraint (\ref{HamiltonianI}) yields $\Delta = -E$.

Our general formulae reduce as desired to the familiar extremal
expressions \eqref{FLOW'II}, \eqref{FLOW'III} when $\Delta=0$, in
which case equation (\ref{Y-equation}) implies that $Y(\varphi,
\phi^a)$ must factor as
\begin{equation} Y(\varphi,
\phi^a)=\e^{\beta\varphi}W(\phi^a)\,.\label{factorization}
\end{equation}
This factorisation property is the main difference between extremal
and non-extremal flow equations.

\section{Existence of a generalised superpotential}\label{s:Existence}

For extremal black hole solutions involving one scalar field a
superpotential always exists \cite{Skenderis:2006jq}: assume that
the extremal solution exists, then equation (\ref{FLOW'II}) defines
the function $W(\tau)$. Since the black hole is supported by a single
scalar $\phi$, and locally we
can always invert $\phi(\tau)$ to $\tau(\phi)$, this
defines $W(\phi)$.\footnote{Having constructed the fake superpotential
$W(\phi)$ for the extremal
solution,
we could then attempt the deformation technique of
\cite{Miller:2006ay}
to obtain the function $Y(\varphi,\phi)$ in the non-extremal case.
This approach,
however, requires the Lagrangian to have certain properties
(see \cite{Miller:2006ay} for details), which do not hold in
general.}
In case of multiple
scalars, the above argument for the existence of \emph{extremal}
flow equations does not apply \cite{Sonner:2007cp}.\footnote{Unless
some complicated
conditions are satisfied, as explained in the case of domain walls in
\cite{Celi:2004st,Sonner:2007cp}.}

For non-extremal solutions carried by an arbitrary number of scalar fields,
the superpotential can be proven to exist under certain conditions
\cite{Perz:2008kh}, generalising the results of \cite{Ceresole:2007wx}.
Since both the `warp factor' $\varphi$ of the $(D+1)$-dimensional
metric and the $(D+1)$-dimensional scalars $\phi^a$ appear on the same
footing in equations
(\ref{FLOWII}) and (\ref{FLOWIII}), we combine them in a vector
$\phi^A$:
\begin{equation}
\phi^A=\{\varphi,\phi^a\}\,.
\end{equation}
In the following section we study a class of theories
with a symmetric moduli space when reduced over one dimension;
their equations of motion are known to be integrable. The
integrability of the effective action allows to explicitly write down the
velocity vector field $f$ on the enlarged scalar manifold in $D$
dimensions
\begin{align}
&\dot{\phi}^A\equiv f^A(\phi,\chi)\,,\\
&\dot{\chi^{\alpha}}\equiv f^{\alpha}(\phi,\chi)\,,
\end{align}
where the $\chi^{\alpha}$ are the scalars descending from the vector
potentials upon dimensional reduction. One can demonstrate that
there are enough `integrals of motion' to fully eliminate the
$\chi^{\alpha}$ in terms of the $\phi^A$, such that one can write
down a velocity field on the original target space in $D+1$
dimensions:
\begin{equation}\label{velocity2}
\dot{\phi}^A=f^A(\phi,\chi(\phi))\,.
\end{equation}

Having obtained the velocity field (\ref{velocity2}) on the moduli
space in $D+1$ dimensions, it suffices to show that the velocity
one-form $f_A$ is locally exact
\begin{equation}\label{curl}
f_A(\phi,\chi(\phi))\equiv \tilde
G_{AB}(\phi)f^B(\phi,\chi(\phi))=\partial_A
Y(\phi)\,,
\end{equation}
where $\tilde G$ is the metric on the scalar manifold in the
$D$-dimensional theory.
A necessary and sufficient condition for this to hold locally is, by
Poincar\'e's lemma, that the one-form is closed
\begin{equation}\label{curl1}
\partial_{[A} f_{B]}=0\,.
\end{equation}

For specific non-supersymmetric solutions it might be very difficult
in practice to find the superpotential $Y$. In spite of this, by
verifying the vanishing curl condition (\ref{curl1}) one can
demonstrate the existence of a gradient flow.\footnote{In some cases
a direct integration turns out to be possible for an \emph{extremal}
ansatz, as in \cite{Hotta:2007wz,Gimon:2007mh}. One can readily check that the
velocity field is irrotational in these examples.} For this reason we
restrict ourselves to those theories that have a symmetric moduli
space after timelike reduction, where we know that $f$ exists.

\section{Black holes and geodesics}\label{s:FreeParticle}

To arrive at an explicit expression
for the velocity field $\dot \phi^A = \{\dot \varphi, \dot
\phi^a\}$ for theories with symmetric moduli spaces after
dimensional reduction \cite{Perz:2008kh}, we first consider
a timelike reduction of the $D+1$-dimensional
theory and then give the necessary background on geodesics on
symmetric spaces.

\subsection{Timelike dimensional reduction}

There is another way to interpret the one-dimensional effective
action given above, first described in the $D+1 = 4$ case in
\cite{Breitenlohner:1987dg}.\footnote{We refer to
\cite{Bergshoeff:2008be} for a recent discussion and application of
this formalism for black holes in symmetric supergravities.}
It is based on the observation that a static solution in $D+1$
dimensions can be dimensionally reduced over time (a Killing
direction) to a Euclidean $D$-dimensional instanton solution.
Because of the assumed spherical symmetry, the resulting instanton
solutions are carried only by the metric and the scalars in $D$
dimensions. We interpret the metric \eqref{metric} as the ansatz for
a dimensional reduction over time. The scalar field equations of
motion are found from the following effective one-dimensional action
\begin{equation}
S= - \frac12 \int \de \tau \,\tilde
G_{ij}\dot{\tilde\phi}^i\dot{\tilde\phi}^j\,,\label{action2}
\end{equation}
which describes the free geodesic motion of a particle in
an enlarged target space of scalar fields $\tilde \phi^i=\{\phi^A, \chi^\alpha\}$,
where the $\phi^A$ contain both the scalars $\phi^a$ of the
$(D,1)$-dimensional theory  and the `warp factor' $\varphi$, and the
$\chi^\alpha$ are axions, consisting of electric potentials (and
magnetic potentials when $D+1=4$ ).\footnote{When going from four to
three dimensions, there is also a scalar dual to the Taub-NUT
vector. Since we only consider static black hole solutions, we
restrict to geodesics for which the NUT charge vanishes.} We always use
the notation $\tilde G$ for the moduli space metric in the reduced
(Euclidean) gravity theory. Note that in this procedure the vectors
(or equivalently, the axions) are not eliminated by their equations
of motion. This action has to be complemented by the Hamiltonian
constraint \cite{Janssen:2007rc} (compare with \eqref{HamiltonianI})
\begin{equation}
\tfrac12\tilde G_{ij}\dot{\tilde\phi}^i
\dot{\tilde\phi}^j\equiv E\,.\label{HamiltonianII}
\end{equation}
Those $D$-dimensional solutions that lift to extremal black holes in
$(D,1)$ dimensions have flat $D$-dimensional geometries, or
equivalently $E=0$, which implies that the geodesic is null: $\tilde
G_{ij}\dot{\tilde\phi}^i \dot{\tilde\phi}^j=0$.

For static, spherically symmetric solutions one can eliminate
the axions $\chi^{\alpha}$ from the action,
since the moduli space metric has the following crucial properties:
$\tilde G_{\alpha A}=0\,, \partial_{\alpha} \tilde G_{ij}=0$. These
identities stem from the fact that the shift
symmetries of the scalars $\phi^\alpha$ in $D+1$ dimensions commute.%
\footnote{In fact, for $D>3$ these properties also hold for
stationary solutions. In $D=3$, the shift symmetries associated with
electric and magnetic charges $q_I,p^I$ no longer commute for
solutions with a non-vanishing NUT-charge. However, the mentioned
properties of the moduli space metric are also valid in $D=3$
upon truncation of the vectorial direction that corresponds to the NUT
charge. Such solutions are spherically symmetric.}

\subsection{Geodesics on symmetric spaces}

Let us now assume that the target space in $D$
dimensions is a symmetric coset space $G/H$, where $G$ is a Lie
group and $H$ some subgroup subject to certain conditions that we
state below. This assumption is always valid for supergravity
theories with more than eight supercharges and for some theories with
less supersymmetry. Nevertheless, our analysis here is independent
of any supersymmetry considerations.

The Lie algebras associated
to $G$ and $H$ are denoted by $\mathfrak{g}$ and $\mathfrak{h}$
respectively. The defining property of a symmetric space $G/H$ is that
there exists a Cartan decomposition
\begin{equation}
\mathfrak{g} = \mathfrak{h} + \mathfrak{f}\,,
\end{equation}
with respect to the Cartan
automorphic involution $\theta$, such that $\theta (\mathfrak{f})= -
\mathfrak{f}$ and $\theta(\mathfrak{h}) = + \mathfrak{h}$.
Take a coset representative $L(\tilde \phi) \in G$. We first
define the group
multiplication from the left, $L\rightarrow gL$, $\forall g\in G$, and
we let the local symmetry act from the right $L\rightarrow Lh \sim L
$, $\forall h \in H$.
From the Cartan involution we can construct the symmetric coset matrix
$M=LL^{\sharp}$, where $\sharp$ is the generalised transpose
\begin{equation}
L^{\sharp}=\exp[-\theta(\log L)]\,.
\end{equation}
The matrix $M$ is invariant under $H$-transformations that act from
the right on $L$.  Under  $G$-transformations from the left, $M$
transforms as $M \rightarrow g M g^{\sharp}$.

With the aid of the matrix $M$ the line element on the space $G/H$
with coordinates $\tilde\phi^i$ can be written as
\begin{equation}
\de s^2 = \tilde G_{ij}\de\tilde\phi^i\de\tilde\phi^j =
-\tfrac12\tr\bigl(\de M \de M^{-1}\bigr)\,.
\end{equation}
A $\emph{local}$ action of $H$ on $L$
from the right and a \emph{global} action of $G$ on $L$ from
the left leave the metric invariant. The latter implies that $G$ is
the isometry group of
$G/H$. The action \eqref{action2} of the dimensionally
reduced theory then describes the geodesic curves on $G/H$ and the
resulting equations of motion are
\begin{equation}\label{eq:scalarEOM}
\tfrac{\de}{\de \tau}(M^{-1}\tfrac{\de}{\de \tau}M)=0\qquad
\Rightarrow\qquad M^{-1}\tfrac{\de}{\de \tau}M=Q\,,
\end{equation}
with the matrix of Noether charges $Q$ being a constant matrix in some
representation of $\mathfrak{g}$. We now see that the geodesic
equations are indeed integrable and their general solution is
\begin{equation}
M(\tau)=M(0) \e^{Q\tau}\,.\label{SOLUTION}
\end{equation}
The affine velocity squared of the geodesic curve is (the dot stands
for ordinary matrix multiplication)
\begin{equation}\label{eq:affinev}
\tilde G_{ij}\dot{\tilde\phi}^i\dot{\tilde\phi}^j =
\tfrac{1}{2}\tr(Q\cdot Q)\,,
\end{equation}
 and coincides with the Hamiltonian constraint \eqref{HamiltonianII}.

An integrable geodesic motion on an $n$-dimensional space is
specified by $2n$ constants: the initial position and velocity
of the geodesic curve. So the geodesic motion on $G/H$ is
specified by \mbox{$2(\dim G-\dim H)$} integration constants. In
eq.~(\ref{SOLUTION}) $M(0)$ contains $(\dim G-\dim H)$
constants corresponding the initial position. The number of
arbitrary constants in $Q$ (the initial velocity) is reduced from
$\dim G$ to $(\dim G-\dim H)$
through the constraint $M^{\sharp}(\tau)=M(\tau)$, which gives
$\theta(Q)=-M(0)^{-1}QM(0)$.

The first-order equation (\ref{eq:scalarEOM})
can be written compactly as $M^{-1}\partial_i M
\dot{\tilde\phi}^i=Q$
or equivalently,
\begin{equation}
\dot{\tilde\phi}^i = \tfrac12\tilde G^{ij} \tr~(M^{-1}\partial_i M
\cdot Q)\,.\label{eq:fo_eqn-general}
\end{equation}
These are only $(\dim G-\dim H)$ equations. After substituting
\eqref{eq:fo_eqn-general} into eq.~\eqref{eq:scalarEOM}, the remaining
$\dim H$ components become non-differential equations. This shows the
power of
\eqref{eq:scalarEOM}: we split the $\dim G$ differential
equations in $M^{-1}\tfrac{\de}{\de \tau}M=Q$ into $(\dim G-\dim H)$
first-order equations and $\dim H$ equations without any derivatives.
In the context of section \ref{s:Existence} these non-differential
equations are precisely what is needed to eliminate the additional
scalars resulting from dimensional reduction, so that we obtain
first-order equations in terms of the scalars in $D+1$ dimensions, as
in eq.~\eqref{velocity2}.

\section{Applications and discussion}\label{s:Discussion}

The gradient flow equations described here descend from a
generalised superpotential and are equally applicable to extremal
(whether supersymmetric or not) as well as non-extremal black holes
(necessarily non-supersymmetric). They naturally encompass previously
known partial results, albeit differ from the form conjectured
in \cite{Andrianopoli:2007gt}.

For theories with scalar manifolds being symmetric
spaces after a timelike dimensional reduction, we give a method
of verifying whether a generalised superpotential exists. It relies
on the fact that the black hole solutions trace out integrable
geodesics on the moduli space of the theory when reduced over time.

We applied these general results to two examples (for details we
refer the reader to \cite{Perz:2008kh}). For a dilatonic
Einstein--Maxwell black hole in four dimensions, where we generalised
the earlier work of \cite{Janssen:2007rc}, we were able to show the
existence of a superpotential even when the procedure of deforming the
extremal solution \cite{Miller:2006ay} cannot be employed.
For the Kaluza--Klein black hole in five dimensions,
which was our second test application, we demonstrated that the
condition for the existence of a generalised superpotential,
which can be viewed as a restriction on the charges, is
nontrivial and independent of extremality.

These findings show that, although it is possible to introduce a
(generalised) superpotential also for non-extremal solutions,
not all black holes of the class discussed, not even all extremal ones,
can be described by a gradient flow, but only those that carry a
specific combination of charges. It would be interesting to investigate
if such charge configurations distinguish themselves through other
physically or mathematically significant properties.

\begin{acknowledgement}
We thank L. Andrianopoli, W. Chemissany, G. Dall'Agata, R. D'Auria,
I. De Baetselier, J. De Rydt, E. Orazi, M. Trigiante and  A. Van
Proeyen for helpful discussions. J.P. is grateful to the organisers of
the 4th RTN Workshop in Varna for the opportunity to present this work.
P.S. has been supported by the German Science Foundation (DFG), T.V.R. by
the G\"oran Gustafsson Foundation, J.P. and B.V. in part by the European
Community's Human Potential Programme under contract
MRTN-CT-2004-005104 `Constituents, fundamental forces and symmetries
of the universe', in part by the FWO-Vlaanderen, project G.0235.05
and in part by the Federal Office for Scientific, Technical and
Cultural Affairs through the `Interuniversity Attraction Poles
Programme -- Belgian Science Policy' P6/11-P\@. B.V. is aspirant
FWO-Vlaanderen.
\end{acknowledgement}

\providecommand{\WileyBibTextsc}{}
\let\textsc\WileyBibTextsc
\providecommand{\othercit}{}
\providecommand{\jr}[1]{#1}
\providecommand{\etal}{~et~al.}

\end{document}